\documentstyle[12pt,fleqn,epsfig]{article}
\begin{document}
\begin{center}

\begin{center}
{\large \bf Onset of Deconfinement in Nucleus-Nucleus Collisions}
\end{center}

\begin{center}
{\large \bf - Past, Present and Future - }
\end{center}

\vspace{1cm}
\begin{center}

Marek Ga\'zdzicki\footnote{E--mail: marek.gazdzicki@cern.ch}\\
Institut f\"ur Kernphysik, Universit\"at Frankfurt, 
Germany and\\
Swietokrzyska Academy, Kielce, Poland 

\end{center}

\vspace{1.5cm}

\begin{minipage}{14cm}
\baselineskip=12pt
\parindent=0.5cm
{\small 
In 2007 Mark I. Gorenstein celebrated his 60th birthday.
This report is dedicated to Mark and it sketches 
the results obtained during the past ten
years of our collaboration and friendship.
They concern 
search for and  study
of the onset of deconfinement in high energy nucleus-nucleus
collisions.
} 

\end{minipage}

\end{center}

\vfill
\noindent
{\footnotesize 
invited talk given at the conference
"New Trends in High Energy Physics", \\
Yalta, Crimea, September 15-22, 2007
}
\newpage

\section{Remarks on the past}

Since the very beginning Mark Gorenstein's
scientific effort is devoted to: \\
- the development of the statistical theory of strong interactions,\\
- the development of statistical models 
  of high energy nucleus-nucleus (A+A) collisions and \\
- the interpretation of the experimental data with the aim to
  uncover properties of strongly interacting matter.

\noindent
This is well illustrated by quoting according to SPIRES Mark's first two
papers from the beginning of~70s:

\vspace{0.1cm}
\noindent
{\it 
1)  Statistical-theory view on the dual-resonance model. \\
M.J. Gorenstein, V.I. Makarov, V.A. Miransky, 
V.P. Shelest, G.M. Zinovjev (BITP, Kiev) .  
Lett.Nuovo Cim.3:347-350,1972. 
}

\vspace{0.1cm}
\noindent
{\it 
2)  A fresh look at the statistical bootstrap model. \\
M. I. Gorenstein, V.A. Miransky, V.P. Shelest, 
G.M. Zinovjev (BITP, Kiev) .  
Published in Phys.Lett.B45:475-477,1973. 
}

\vspace{0.1cm}
\noindent
his middle papers from the 90s,
which mark a start of his travels abroad and a collaboration
with physicists from all over the world:

\vspace{0.1cm}
\noindent
{\it 
103)  Phase-transitions in nuclear matter: Deconfinement 
of quarks and gluons? \\
D.H. Rischke, M. I. Gorenstein, A. Schafer, 
H. Stoecker, W. Greiner (Frankfurt U.). \\
Given at 20th International Workshop on Gross Properties of Nuclei and Nuclear Excitations, 
Hirschegg, Austria, 20-25 Jan 1992. 
}

\vspace{0.1cm}
\noindent
{\it 
104)  Excluded volume effect and the quark - hadron phase transition. \\
J. Cleymans (Cape Town U. \& Bielefeld U.), 
M. I. Gorenstein (Frankfurt U. \& BITP, Kiev) ,
J. Stalnacke, E. Suhonen (Oulu U.). 
Published in Phys.Scripta 48:277-280,1993.
}

\vspace{0.1cm}
\noindent
and finally his most recent two papers:

\vspace{0.1cm}
\noindent
{\it 
208)  Fluctuations in Statistical Models.
M. I. Gorenstein . Sep 2007. Temporary entry
e-Print: arXiv:0709.1428 [nucl-th].
}

\vspace{0.1cm}
\noindent
{\it 
209)  Bose-Einstein Condensation of Pions.
V. Begun, M. I. Gorenstein . Sep 2007. Temporary entry
e-Print: arXiv:0709.1434 [hep-ph].
}

\vspace{0.1cm}
\noindent
The unique property of Mark's work is 
the clarity and simplicity
in attempts to understand
the objective properties of strongly interacting matter
at high energy densities.
Of particular interest for Mark are the questions: \\
- What are the phases of strongly interacting matter? \\
- Does the hypothetical high energy density state of strongly
  interacting matter, Quark Gluon Plasma (QGP), exist?  and \\
- How do the transitions between the phases of strongly 
  interacting matter look like?

\begin{figure}[ht!]
\epsfig{file=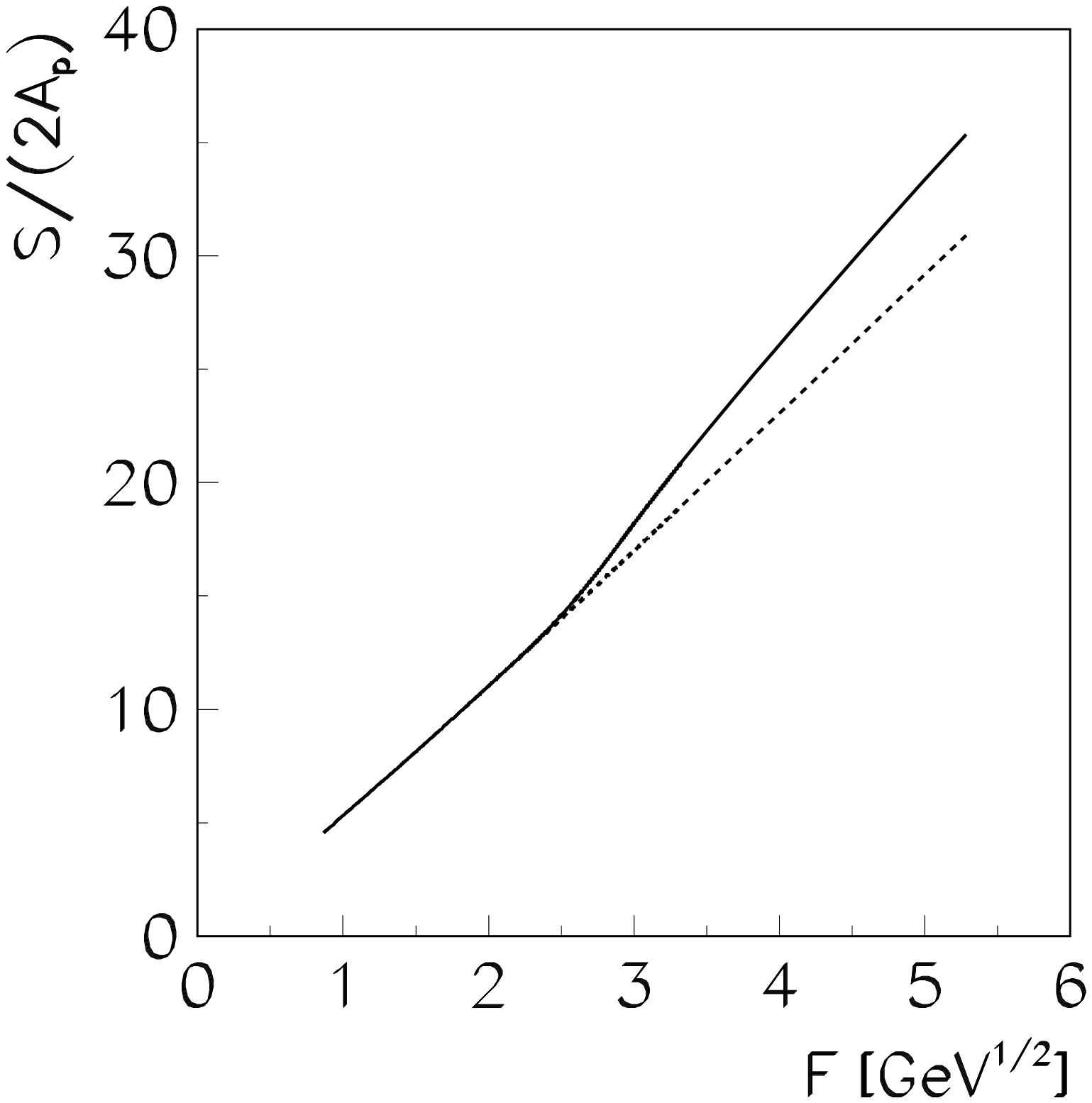, width=0.46\textwidth}
\epsfig{file=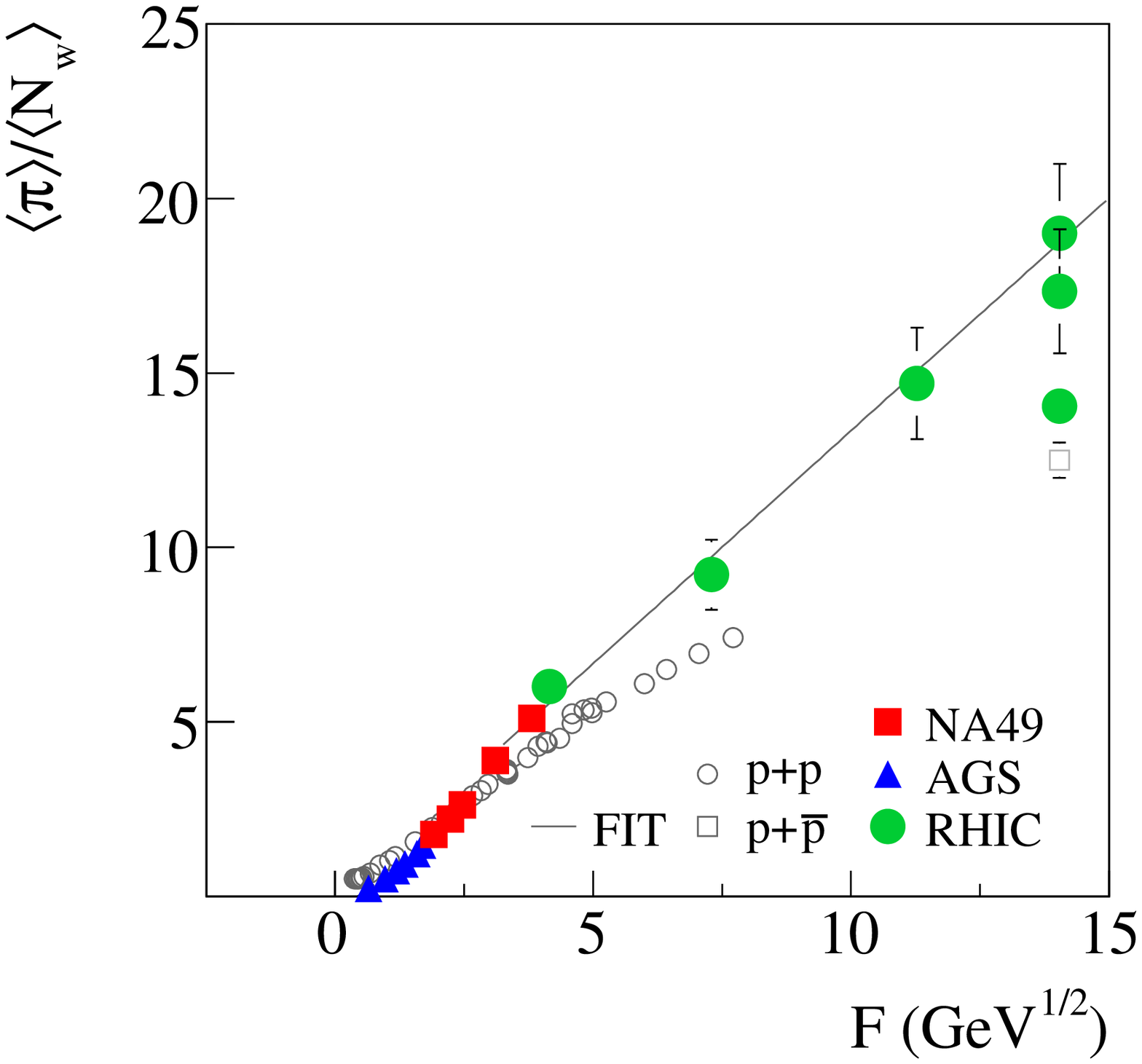, width=0.48\textwidth}
\caption{
Left: 
The energy dependence of the early stage entropy
calculated with SMES~\cite{GaGo}. 
A change of the slope at $F \approx 2.25$~GeV$^{1/2}$
($\approx 30A$~GeV)
indicates the onset of deconfinement.
Right:
Energy dependence of the mean pion multiplicity per wounded
nucleon measured in central Pb+Pb and Au+Au
collisions (full symbols), compared to the corresponding results
from $p+p$ reactions (open circles)~\cite{2030}.
}
\label{pions}
\end{figure}

In the mid-90s, when I met Mark  for the first time during his 
stay at the University of Frankfurt, two predicted signals \cite{rafelski,satz}
of QGP creation in A+A collisions
were observed in central A+A collisions at the top
SPS energy, 158$A$ GeV. These were: \\
- the enhanced production of strange hadrons \cite{na35} and \\
- the suppressed production of J/$\psi$ mesons \cite{na38}. \\
However, soon after it was realized  that these signals
are not specific for Quark Gluon Plasma creation.
In particular,
the strangeness enhancement increases with decreasing collision
energy \cite{GaGoRo} and production properties 
of J/$\psi$ mesons can be
explained assuming that it is produced together 
with other hadrons \cite{GaGopsi}, e.g. via statistical
hadronization process.
On the other hand
a comparison of the first results on nucleus-nucleus collisions
at the top BNL AGS and
CERN SPS energies \cite{GaRo}
suggested  an anomaly located
between these two energies which might signal the onset
of deconfinement at the early stage of nucleus-nucleus collisions.
In numerous non-ending discussions we tried to find
a consistent interpretation of the experimental results.
It soon became obvious for us that an elegant explanation 
can be formulated provided that
the commonly accepted paradigm \cite{rafelski} 
concerning a long-lasting dynamical stage of 
strange quark or hadron production in A+A collisions is
replaced by an assumption of a very fast equilibration of the
matter created at the early stage of collisions.

A model based on the later assumption was proposed by Mark and myself
in 1997 \cite{GaGo} and we called it the Statistical Model of
the Early Stage (SMES).
Within this model the collision energy crosses
the threshold energy needed for the creation of the mixed 
phase at the low SPS energies. The model predicted 
several contra-intuitive (within the old paradigm)
signals of the onset of deconfinement.
Before the final publication in Acta Physica Polonica in 1999
the paper was rejected by Physical Review C, Journal of
High Energy Physics and Journal of Physics G.
Now it has more than 150 citations.
This long publication history illustrates well how
controversial were, and probably still are, for many
our colleagues the assumptions and the predictions of SMES. 

\begin{figure}[ht!]
\epsfig{file=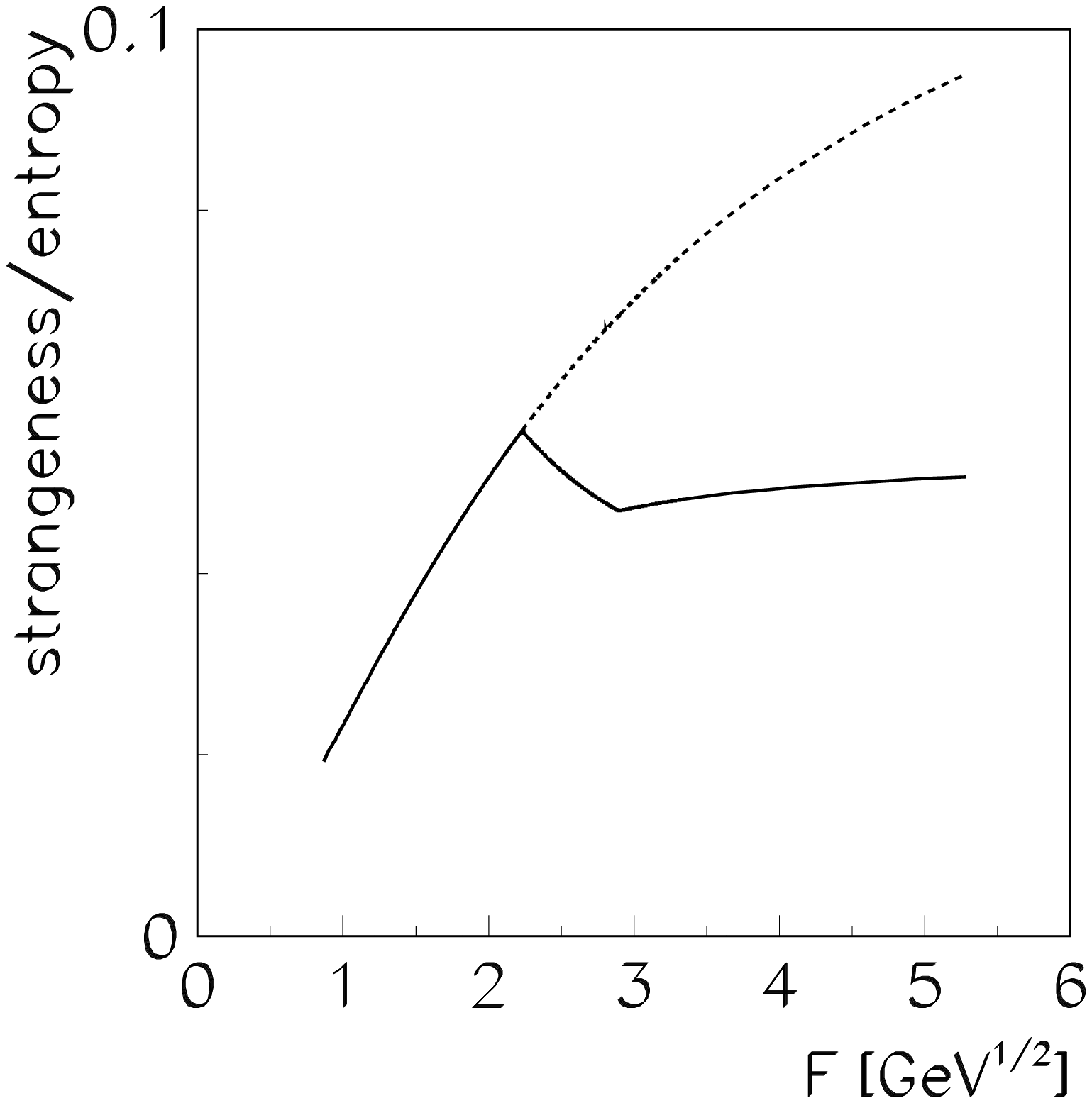, width=0.46\textwidth}
\epsfig{file=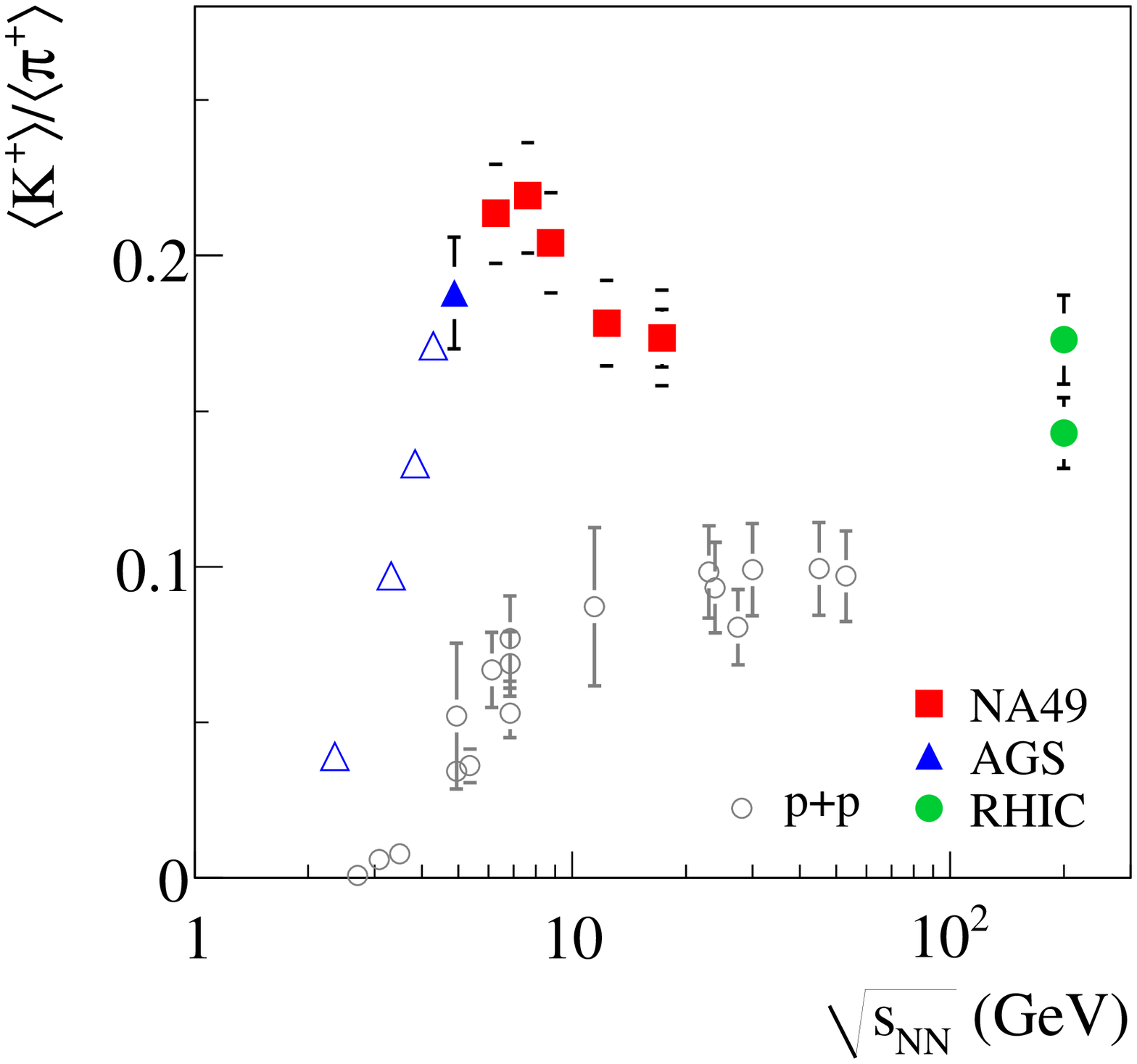, width=0.48\textwidth}
\caption{
Left: 
The energy dependence of the strangeness to entropy ratio
calculated with SMES (solid line). 
A maximum   at $F \approx 2.25$~GeV$^{1/2}$
($\approx 30A$~GeV)
indicates the onset of deconfinement~\cite{GaGo}.
Right:
Energy dependence of the $\langle K^+ \rangle / \langle \pi^+ \rangle$ 
ratio
measured in central Pb+Pb and Au+Au collisions
(full symbols) compared to the corresponding results from $p+p$
reactions (open circles)~\cite{2030}.
}
\label{strangeness}
\end{figure}

In spite of the controversy and thanks to a strong support
of Peter Seyboth, the NA49 spokesperson, the NA49
Collaboration proposed in 1997 the so-called energy scan program,
at the CERN SPS.
The goal of this effort was to check the SMES predictions.

The  program  started with two runs
where data on central Pb+Pb collisions at 40$A$ and 80$A$~GeV were
recorded in 1999 and 2000. Data at the top SPS energy of 158$A$~GeV  had
already been taken in previous SPS runs.  The analysis of these runs
was published in~\cite{Afanasiev:2002mx} and the results confirmed the
predictions of~\cite{GaGo}. This finding motivated an extension of the
energy scan to the lower SPS energies of 30$A$ and 20$A$~GeV  which was
completed in 2002 whereas the final results were published 
recently \cite{2030}.

\section{Remarks on the Present}

The final NA49 results from the energy scan program together
with reach  measurements at the AGS and RHIC energies
for the first time allow to check SMES predictions over a
broad energy range. 
This section is devoted to this subject.

The SMES model is based on the
assumption that the system created at the early stage (be it confined
matter or a QGP) is in equilibrium and that a transition from a
reaction with purely confined matter to a reaction with a QGP at the
early stage occurs when the transition temperature $T_C$ is reached.
For $T_C$ values of 170--200~MeV the transition region extends from 15$A$
to 60$A$~GeV~\cite{GaGo}.  Assuming the generalized Fermi-Landau
conditions~\cite{GaGo,Fe:50} for the early stage of
nucleus-nucleus collisions and a proportionality of the pion
multiplicity to the early stage entropy, the ratio 
mean pion multiplicity to the number of wounded nucleons 
($ \langle \pi \rangle /\langle N_W \rangle$)
increases linearly with the Fermi 
collision energy measure 
($ F \equiv (\sqrt{s_{NN}} - 2 m_N)^{3/4} /
   \sqrt{s_{NN}}^{1/4} $) 
outside the transition region.
The SMES prediction on the energy dependence of the early stage
entropy is shown in Fig.~\ref{pions} (left).
The slope parameter is proportional to
$g^{1/4}$~\cite{GaGo}, where $g$ is the effective number of internal
degrees of freedom at the early stage.  In the transition region a
steepening of the pion energy dependence is due to the activation of a
large number of partonic degrees of freedom.  This is, in fact,
observed in the data on central Pb+Pb and Au+Au collisions, where the
steepening starts at about 20$A$~GeV, as shown in Fig.~\ref{pions} (right).
The linear dependence of $ \angle \pi \rangle / \langle N_W \rangle$ 
on $F$ is approximately
obeyed by the data at lower and higher energies (including RHIC).  An
increase of the slope by a factor of about 1.3 is measured, which
corresponds to an increase of the effective number of internal degrees
of freedom by a factor of 1.3$^4$ $\cong$ 3, within the
SMES.

The $\langle K^+ \rangle / \langle \pi^+ \rangle$ ratio 
shown in Fig.~\ref{strangeness} (right)  
is approximately
proportional to the total strangeness to entropy ratio which in the
SMES model is assumed to be preserved from the early stage till
freeze-out.  At the low collision energies the strangeness to entropy
ratio increases steeply with the collision energy, due to the low
temperature at the early stage ($T < T_C$) and the high mass of the
strangeness carriers in the confined state (the kaon mass,
for instance, is 500~MeV).  When the transition to a QGP is crossed
($T > T_C$), the mass of the strangeness carriers is significantly
reduced to the strange quark mass of about 100~MeV.  Due to the low
mass $m < T$, the strangeness yield becomes (approximately)
proportional to the entropy, and the strangeness to entropy (or pion)
ratio is independent of energy.  This leads to a decrease in the
energy dependence from the larger value for confined matter at $T_C$ to
the QGP value. 
Thus the  non-monotonic energy dependence of
the strangeness to entropy ratio is followed by a saturation at the
QGP value. 
The predicted behavior is shown in Fig.~\ref{strangeness} (left).
Such anomalous energy dependence can indeed be seen
in~Fig.~\ref{strangeness} (right) and is, within the SMES, 
a direct consequence
of the onset of deconfinement taking place at about 30$A$~GeV.

\begin{figure}[ht!]
\epsfig{file=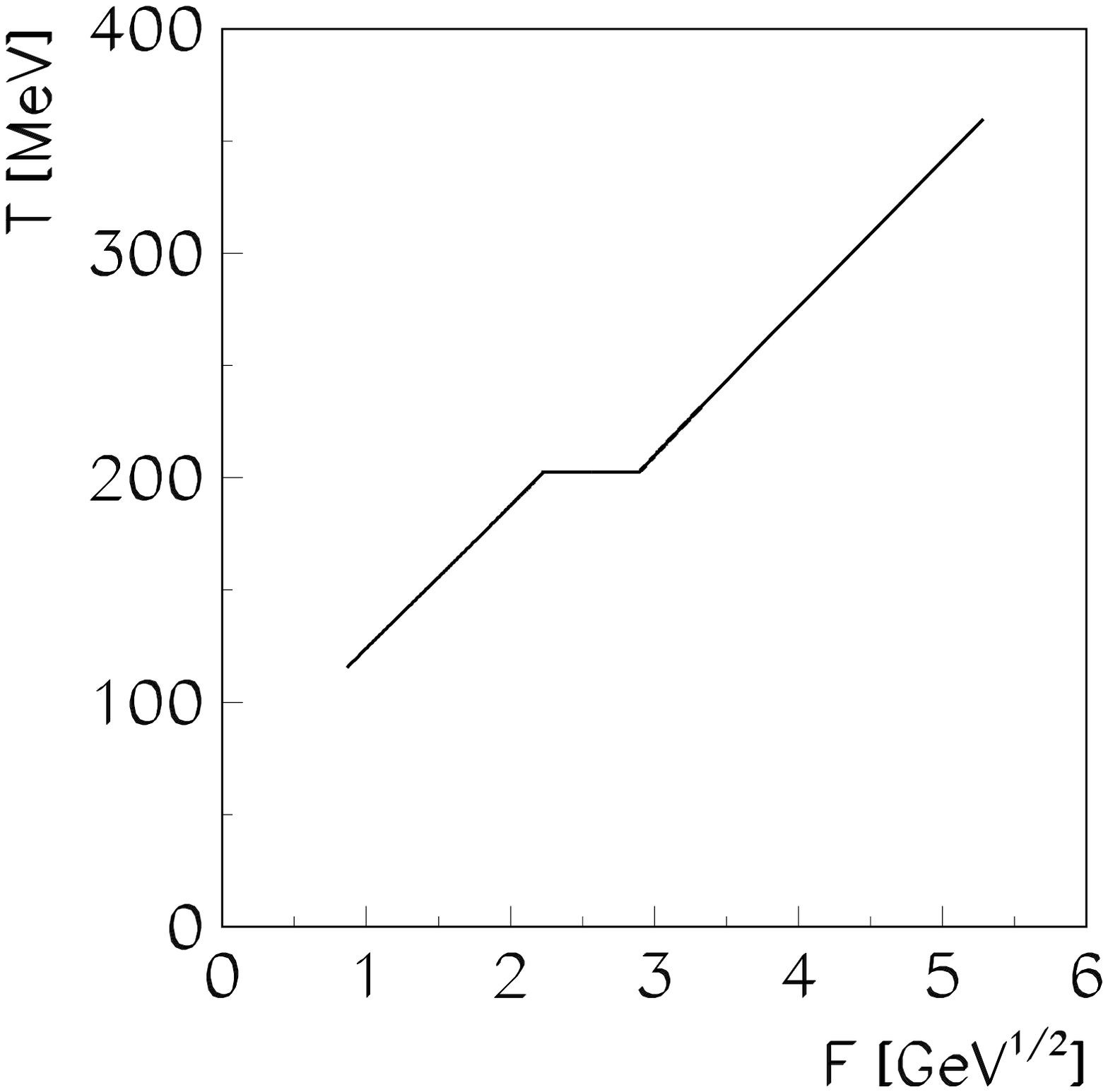, width=0.46\textwidth}
\epsfig{file=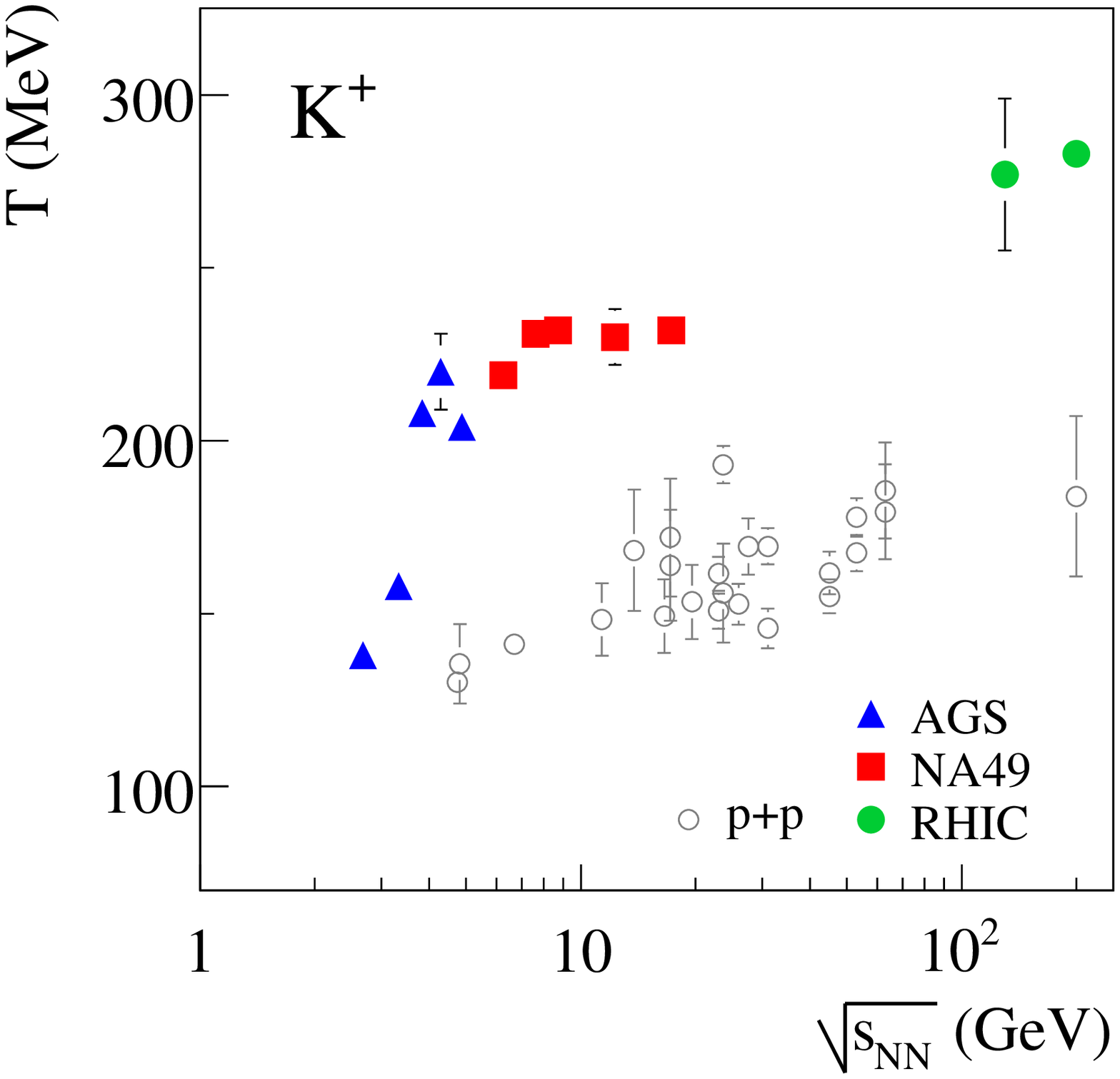, width=0.48\textwidth}
\caption{
Left: 
The energy dependence of the early stage temperature
calculated with SMES (solid line). 
A change of the slope  at $F \approx 2.25$~GeV$^{1/2}$
($\approx 30A$~GeV)
indicates the onset of deconfinement~\cite{GaGo}.
Right:
Energy dependence of the inverse slope parameter of the transverse
mass spectra for $K^+$ mesons 
measured in central Pb+Pb and Au+Au collisions
(full symbols) compared to the corresponding results from $p+p$
reactions (open circles)~\cite{2030}.
}
\label{slope}
\end{figure}

In the mixed phase region the early stage pressure and temperature are
independent of the energy density~\cite{GaGo}. Within SMES this
causes a step like dependence of the pressure and temperature 
on the collision energy, see
Fig.~\ref{slope} (left) for illustration.
Consequently, this should lead to the weakening
of the increase with energy of the inverse slope parameter $T$ or,
equivalently, the mean transverse mass $\langle m_T \rangle$ 
in the SPS energy
range~\cite{Gorenstein:2003cu}.  This qualitative prediction is
confirmed by the results shown in Fig.~\ref{slope} (right).

\section{Remarks on the Future}

Observation of the signals of the onset of deconfinement
in central Pb+Pb collisions at the low SPS energies 
leads to new experimental and theoretical questions.
Among experimental challenges are:\\
- A confirmation of the NA49 results, \\
- A study of the system size dependence of the observed effects,\\
- A search for signals in fluctuations and correlations.\\
These will be addressed by
several experiments  planned in the CERN SPS energy range:
NA61 at the CERN SPS \cite{na61}, 
STAR and PHENIX at the BNL RHIC \cite{rhic},
MPD at JINR NICA \cite{mpd} and CBM at FAIR SIS-300 \cite{cbm}.

\begin{figure}[ht!]
\begin{center}
\epsfig{file=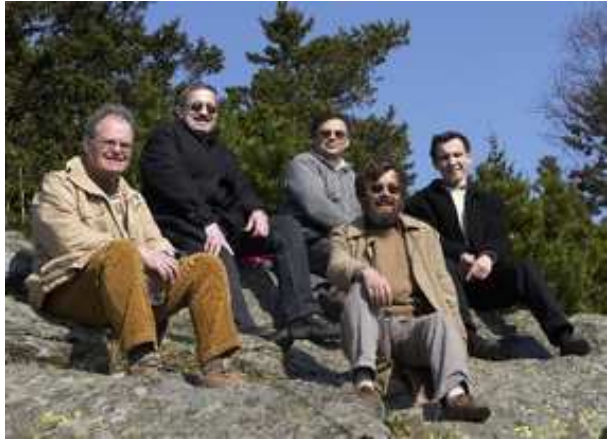, width=0.50\textwidth}
\end{center}
\caption{
Mark (the second from the left) with his friends 
and collaborators in Bergen,~2005.
}
\label{mark}
\end{figure}

Among  new theoretical questions are:\\
- What is the mechanism of the early stage equilibration? \\
- What is the structure of the transition line (cross-over,
  the critical point, 1st order phase transition, ...)? \\
- How can this structure  be established experimentally? \\
In particular, the last two questions attract Mark's attention.

Recently, Mark returned to the quark-gluon bag model
formulated by him in the 80s \cite{qgb} and together 
with his collaborators
he suggested an unusual structure of the transition line
\cite{line}. I hope Mark will continue this development
and will try to predict the resulting experimental signals.

Furthermore,
he initiated \cite{fluct} and now leads with his collaborators
(see about papers of Mark et al.
on this subject)  the
pioneering study of fluctuations in relativistic gases.
This effort resulted in predictions of many new effects
related to the conservation laws, quantum statistics
and the finite volume of hadrons. 
These predictions define a base-line with respect to which effects
related to the onset of deconfinement 
\cite{fluct_ood} and the critical point
can be looked for.

\vspace{0.5cm}
Let me finish by a simple wish:\\
\begin{center}
{\bf Dear Mark, just continue ...}
\end{center}

\vfill
\noindent
{\small
I would like to thank the organizers of the conference 
New Trends in High Energy Physics (Yalta, Crimea, September 15-22, 2007)
for the interesting and inspiring meeting.
This work was supported by the Virtual Institute VI-146
of Helmholtz Gemeinschaft, Germany.}

\end{document}